\documentclass[twocolumn ,amsmath, amssymb, longbibliography,10pt, citeautoscript,superscriptaddress ]{revtex4-2}
\usepackage{amsmath}
\usepackage{graphicx}
\usepackage{dcolumn}
\usepackage{natbib}

\usepackage{amsfonts}
\usepackage{latexsym}
\usepackage{graphicx}
\usepackage{color}
\usepackage{bm}
\usepackage{float}

\definecolor{Blue}{rgb}{0,0,1}

\usepackage{hyperref} 
\usepackage[all]{hypcap}
\hypersetup{pdfauthor = {First Author},pdftitle = {}}

\begin{document}
\title{Nonlocal action at a distance also acts in the past}
\author{Mandip Singh}
\email{mandip@iisermohali.ac.in}
\affiliation{Department of Physical Sciences,
\\Indian Institute of Science Education and Research Mohali, Sector-81, S.A.S. Nagar, Mohali, 140306, India.}

\begin{abstract}
 The nonlocal action of a measurement performed on a quantum entangled particle can determine the quantum state of a distant entangled particle instantly. Since the relativistic simultaneity of events is frame dependent therefore, a physically valid question arises, does the nonlocal action at a distance determine the quantum state of the distant particle in its past, when quantum state collapse is observed from a different inertial frame of reference? From the relativity of simultaneity, Lorentz invariance of the quantum entangled state of photons under consideration and the validity of Bell's theorem in all inertial frames of reference, it is demonstrated that the measurement-induced collapse of a quantum entangled state also determines the quantum state of a distant photon in its past, provided one of the photons is located in the future.  The past and future are defined with respect to the time of quantum state collapse in the rest frame of a quantum state measurement device.
\end{abstract}
\maketitle
\section{Introduction}
Einstein Podolsky and Rosen, in their seminal paper, argued that quantum mechanics is inconsistent with the principle of locality and the notion of reality \cite{epr}. According to the principle of locality, any type of measurement performed on a particle cannot instantly influence another distant particle, \emph{i.e.} a faster than light influence at a distant location is not possible. Furthermore, the classical notion of reality signifies that all observables have definite values even prior to a measurement, which can be measured without disturbing the system. However, according to the Copenhagen interpretation of quantum mechanics, the wavefunction contains a complete information of the system by means of which some incompatible observables values are indefinite prior to a measurement. Thus, a measurement is necessary to obtain a definite value of the observable, where a measured value and corresponding quantum state represent a measurement outcome. The measurement outcome can vary randomly in repeated measurements. The action of a measurement disturbs the wavefunction of the quantum system which is known as wavefunction collapse. Moreover, if two particles are prepared in a known maximally entangled quantum state then a single projective measurement on one particle can immediately determine the quantum state of the distant particle from a measurement outcome of the collapsed quantum state. This effect is known as the action at a distance, where the quantum state of each particle was completely undefined prior to a measurement because of their quantum entanglement. Various theories based on local hidden variables are proposed to explain the predictions of quantum mechanics and experimental observations \cite{Shimony2009}. To test whether the local hidden variable models or quantum mechanics are correct, J. Bell introduced an inequality that cannot be violated if a local hidden variable model is correct \cite{bell_original, bell_book, chsh_th}. The correlations predicted by quantum mechanics are stronger and Bell's inequality is violated in different experiments \cite{clauser_bell, bell_aspect_eprb, bell_aspect, chsh_expt, test_thompson, shih_eprb, epr-atom_haroche, bell_mandle} and also under the strict condition of locality \cite{bell_weihs} \emph{i.e.} where distant particles separated by a spacelike interval are measured independently and simultaneously. Various experiments of loophole-free Bell's inequality violation are in agreement with quantum mechanics \cite{bell_elecrons, bell_atom, bell_remote_qubits, bell_phase_qubits, bell_without_fairsam, dt_lh_free_nonlonality, bell_nonlocality_rev, det_loophole, bell_ions_photons, bell_4km_fiber, bell_10km_gisin, long_144}.

 According to the principle of quantum superposition, a single quantum particle can be prepared in different quantum states at the same instant of time. If any projective measurement is performed to measure components of a quantum superposition state, then the quantum superposition state collapses randomly to one of its components. The quantum state collapse is supposed to happen instantly, even for a single particle. It is also demonstrated experimentally that a quantum superposition state collapses almost instantly even when the quantum superposed states are separated far apart \cite{speed_hugo, speed_pan, speed_nicrosini}. However, the notion of the same instant of time or simultaneity changes in different inertial frames of reference moving with a relativistic speed \emph{w.r.t} each other \cite{Einstein_SR}. A region of spacetime at which a quantum measurement device performs a measurement is defined as a collapse event. Since the nonlocal state collapse of a quantum state is supposed to happen instantly in space therefore, quantum state remains collapsed in the future of the quantum state collapse event in the rest frame of the quantum state measurement device. The past and future are defined \emph{w.r.t} the time of the quantum state collapse event in this frame. Since simultaneity is relative therefore, a different relativistically moving inertial frame of reference perceives the quantum state collapse of spatially separated particles at different spacetime locations.  A relativistically moving frame of reference perceives two particles in a different way at the same instant of time, where one particles exists in the past and the other particle exists in the future of the quantum state collapse event in the rest frame of a quantum state measurement device. Since a particle existing in the future has been measured and disentangled at the quantum state collapse event.
Therefore, an important question arises here, does it imply that the collapse of a quantum entangled state also influences the quantum state of the distant unmeasured particle in its past? This question is analyzed in this paper, and it is shown that this is true provided one of the particles is located in the future of the quantum state collapse event.

This paper analyzes the above question with spacetime diagrams of different inertial frames of reference moving with a relativistic speed \emph{w.r.t.} each other without rotation.  Two maximally polarization entangled photons propagate in the opposite direction and one of them is measured to determine its polarization quantum state, where the polarization entangled state under consideration is Lorentz invariant. This single photon measurement collapses the quantum entangled state and  polarization quantum state of the distant photon is immediately determined without making any interaction with it. Where future refers to  the time zone after the quantum state collapse event and  past refers to the time zone before the quantum state collapse event \emph{w.r.t}  a frame of reference in which a quantum measurement device is at rest. The quantum state collapse process is analyzed in different relativistic inertial frames of reference by considering the relative simultaneity, Lorentz invariance of the quantum entangled state under consideration and the validity of Bell's theorem in all inertial frames of reference.

\section{Quantum entanglement of photons in different relativistic inertial frames}
Consider an inertial frame of reference $S'$ moving with uniform velocity $v$ along $x$-axis \emph{w.r.t} an inertial frame of reference $S$. Each frame of reference is associated with a flat spacetime and events are represented by cartesian coordinate systems comprising of coordinates $(x,y,z, ct)$ and $(x',y',z', ct')$ corresponding to frames $S$ and $S'$, respectively. The spacetime events, from one frame to another, are related by the Lorentz transformations; $x'=\gamma(x-vt)$, $y'=y$, $z'=z$, $t'=\gamma(t-v x/c^{2})$, where $c$ is the speed of light and $\gamma=1/(1-v^{2}/c^{2})^{1/2}$. Consider a pair of polarization entangled photons of same frequency $\nu$ with total quantum state $|\psi\rangle_{T}=|p_{1x}\rangle_{1}|p_{2x}\rangle_{2}\otimes\frac{1}{\sqrt{2}}(|\hat{z}\rangle_{1}|\hat{y}\rangle_{2}-|\hat{y}\rangle_{1}|\hat{z}\rangle_{2})$ in frame $S$ and propagating in the opposite direction along $x$-axis such that photon-$1$ propagation along $x$-direction.  Where $|\hat{y}\rangle_{j}$ and $|\hat{z}\rangle_{j}$  represent linear polarization quantum states of photon-$j$ along the direction of unit vectors $\hat{y}$ and $\hat{z}$, respectively. In addition, $|p_{jx}\rangle_{j}$ is the eigen state of $x$-component of momentum operator for photon-$j$ in frame $S$. Therefore, photons  are completely delocalised along $x$-axis.  Time evolution of the quantum state is $|\psi;t\rangle_{T}=e^{-i(|p_{1x}|+|p_{2x}|)c t/\hslash}|\psi;t=0\rangle_{T}$, where $|p_{jx}|= E_{j}/c$ is the magnitude of momentum of photon-$j$ along $x$-axis and $E_{j}=h\nu_{j}$ is energy of photon-$j$ in frame $S$. Frequencies of both photons are same in frame $S$ \emph{i.e.} $\nu_{1}=\nu_{2}$.
However,  according to the uncertainty principle, each photon can be localised in space at a given instant time if the spatial extension of its spatial wavefunction is reduced by increasing the extension of its momentum wavefunction. 
Each photon wavepacket is represented by quantum superposition of its momentum eigen states therefore, total quantum state can be written as
\begin{multline}
\label{2}
  |\alpha;t\rangle_{T}=\int^{\infty}_{-\infty}\int^{\infty}_{-\infty}\psi_{1}(p_{1x})\psi_{2}(p_{2x})e^{-i(|p_{1x}|+|p_{2x}|)c t/\hslash}\\|p_{1x}\rangle_{1}|p_{2x}\rangle_{2}\otimes\frac{1}{\sqrt{2}}(|\hat{z}\rangle_{1}|\hat{y}\rangle_{2}-|\hat{y}\rangle_{1}|\hat{z}\rangle_{2})\mathrm{d}p_{1x}\mathrm{d}p_{2x}
\end{multline}
This quantum state can be expressed in position-space wavefunctions by evaluating ${_{2}\langle} x_{2}|{_{1}\langle} x_{1}|\alpha;t\rangle_{T}$, where $|x_{j}\rangle_{j}$ is the $x$-position operator eigen state of photon-$j$. Therefore,
\begin{multline}
\label{eq3}
 |\Psi;t\rangle_{T} ={_{2}\langle} x_{2}|{_{1}\langle} x_{1}|\alpha;t\rangle_{T}\\=\frac{1}{2\pi\hslash}\int^{\infty}_{-\infty}\psi_{1}(p_{1x})e^{\frac{i (p_{1x}x_{1}-|p_{1x}|c t)}{\hslash}}\mathrm{d}p_{1x}\\\int^{\infty}_{-\infty}\psi_{2}(p_{2x})e^{\frac{i (p_{2x}x_{2}-|p_{2x}|c t)}{\hslash}}\mathrm{d}p_{2x}\\\otimes\frac{1}{\sqrt{2}}(|\hat{z}\rangle_{1}|\hat{y}\rangle_{2}-|\hat{y}\rangle_{1}|\hat{z}\rangle_{2})
\end{multline}
Which is succinctly written as, $|\Psi;t\rangle_{T}= \psi_{1}(x_{1},t)\psi_{2}(x_{2},t)\otimes\frac{1}{\sqrt{2}}(|\hat{z}\rangle_{1}|\hat{y}\rangle_{2}-|\hat{y}\rangle_{1}|\hat{z}\rangle_{2})$,
where $\psi_{1}(x_{1},t)=\frac{1}{\sqrt{2\pi\hslash}}\int^{\infty}_{-\infty}\psi_{1}(p_{1x})e^{i (p_{1x}x_{1}-|p_{1x}|c t)/\hslash}\mathrm{d}p_{1x}$ and $\psi_{2}(x_{2},t)= \frac{1}{\sqrt{2\pi\hslash}}\int^{\infty}_{-\infty}\psi_{2}(p_{2x})e^{i (p_{2x}x_{2}-|p_{2x}|c t)/\hslash}\mathrm{d}p_{2x}$ are time dependent position-space wavefunctions of photon-$1$ and photon-$2$, repectively.
These position wavefunctions can be the Gaussian wavepackets representing photons propagating in the opposite direction. Total quantum state $|\Psi;t\rangle_{T}$ of photons in frame $S$ is a product to their spatial  wavefunctions and their polarization entangled quantum state, where the polarization entangled state is time independent.

In frame $S$, photons have same energy however, in frame $S'$ photon-$1$ has lower energy than photon-$2$ due to the relativistic Doppler shift. In free space, the electromagnetic field of a photon is orthogonal to its direction of propagation in all inertial frames of reference. Furthermore, $y'$ and $z'$ spatial coordinates remain unaffected by the Lorentz transformations. As a consequence of Lorentz transformations, energy of photons is not same in frame $S'$. From the Lorentz transformations of momentum and energy for frames $S$ to $S'$ \emph{i.e.} $p_{x}'=\gamma(p_{x}-vE)$, $p'_{y}=p_{y}$, $p'_{z}=p_{z}$, $E'=\gamma(E-v p_{x}/c^{2})$, the quantum state Eq.~\ref{eq3} can be transformed to frame $S'$.
 By using the Lorentz invariant quantity $p_{x}x-Et=p'_{x}x'-E't'$ and orthogonality properly of photon electromagnetic field in all inertial frames that leads to $|\hat{y}\rangle_{j}\rightarrow|\hat{y}'\rangle_{j}$ and $|\hat{z}\rangle_{j}\rightarrow|\hat{z}'\rangle_{j}$, the quantum state  Eq.~\ref{eq3} in frame $S'$ can be written as
\begin{multline}
\label{eq4}
 |\Psi';t'\rangle_{T} =\frac{1}{2\pi\hslash}\int^{\infty}_{-\infty}\psi'_{1}(p'_{1x})e^{\frac{i (p'_{1x}x'_{1}-|p'_{1x}|c t')}{\hslash}}\mathrm{d}p'_{1x}\\\int^{\infty}_{-\infty}\psi'_{2}(p'_{2x})e^{\frac{i (p'_{2x}x'_{2}-|p'_{2x}|c t')}{\hslash}}\mathrm{d}p'_{2x}\\\otimes\frac{1}{\sqrt{2}}(|\hat{z}'\rangle_{1}|\hat{y}'\rangle_{2}-|\hat{y}'\rangle_{1}|\hat{z}'\rangle_{2})
\end{multline}
Which is succinctly written as, $|\Psi';t'\rangle_{T}= \psi'_{1}(x'_{1},t')\psi'_{2}(x'_{2},t')\otimes\frac{1}{\sqrt{2}}(|\hat{z}'\rangle_{1}|\hat{y}'\rangle_{2}-|\hat{y}'\rangle_{1}|\hat{z}'\rangle_{2})$,
where $\psi'_{1}(x'_{1},t')=\frac{1}{\sqrt{2\pi\hslash}}\int^{\infty}_{-\infty}\psi'_{1}(p'_{1x})e^{i (p'_{1x}x_{1}-|p'_{1x}|c t')/\hslash}\mathrm{d}p'_{1x}$ and $\psi'_{2}(x'_{2},t')= \frac{1}{\sqrt{2\pi\hslash}}\int^{\infty}_{-\infty}\psi'_{2}(p'_{2x})e^{i (p'_{2x}x'_{2}-|p'_{2x}|c t')/\hslash}\mathrm{d}p'_{2x}$ are time dependent position-space wavefunctions of photons in frame $S'$. In frame $S'$, $|\Psi';t'\rangle_{T}$ represents a quantum state of photons along the line of simultaneity at time $t'$. However, the polarization entangled state of photons is time independent and has same form in all inertial frames of reference. Now quantum state collapse can be analyzed in any inertial frame of reference as described further.

\section{Bell's inequality in different relativistic inertial frames}
Bell's inequality for  different inertial observers is presented in this section before proceeding to the effect of collapse of a quantum entangled state in different inertial frames of reference. First, consider quantum correlation measurements in frame $S$. For polarization entangled state $|\Psi;t\rangle_{E}=(|\hat{z}\rangle_{1}|\hat{y}\rangle_{2}-|\hat{y}\rangle_{1}|\hat{z}\rangle_{2})/\sqrt{2}$ given in Eq.~\ref{eq3},  a measurement of operator $\hat{\sigma}^{(j)}_{3}$ produces $\langle\hat{\sigma}^{(1)}_{3} \rangle ={_{E}\langle}\Psi;t|\hat{\sigma}^{(1)}_{3} \otimes \hat{I}^{(2)} |\Psi;t\rangle_{E}$ and $\langle \hat{\sigma}^{(2)}_{3}\rangle={_{E}\langle}\Psi;t|\hat{I}^{(1)} \otimes \hat{\sigma}^{(2)}_{3}|\Psi;t\rangle_{E}$ equal to zero, where $\hat{\sigma}^{(j)}_{3}\doteq(|\hat{z}\rangle\langle \hat{z}|-|\hat{y}\rangle\langle \hat{y}|)_{j}$ is an operator and $\hat{I}^{(j)}$ is an identity operator in the polarization basis of photon-$j$.  A quantum state measurement device measures $\hat{\sigma}^{(1)}_{3}$ and location of photon-$1$, which further determines the polarization state and location of photon-$2$ along the line of simultaneity in frame $S$. Similarly, an expectation value of single photon operators $\hat{\sigma}^{(j)}_{1}\doteq(|\hat{z}\rangle\langle \hat{y}|+|\hat{y}\rangle\langle \hat{z}|)_{j}$, $\hat{\sigma}^{(j)}_{2}\doteq i(-|\hat{z}\rangle\langle \hat{y}|+|\hat{y}\rangle\langle \hat{z}|)_{j}$ is zero. In contrast, an operator  $\hat{\sigma}^{(1)}_{3}\otimes\hat{\sigma}^{(2)}_{3}$ can be measured without collapsing $|\Psi;t\rangle_{E}$ since polarization entangled state is an eigen state of this operator such that $\hat{\sigma}^{(1)}_{3}\otimes\hat{\sigma}^{(2)}_{3}|\Psi;t\rangle_{E}=-|\Psi;t\rangle_{E}$, which can be measured by Bell state analysis if photons overlap.
For spatially separated photons, a  single photon operator measurement is performed on any photon which collapses the entangled quantum state. To measure $\hat{\sigma}^{(1)}_{3}\otimes\hat{\sigma}^{(2)}_{3}$ by the quantum state collapse, record the measurement outcomes of $\hat{\sigma}^{(1)}_{3}$ and $\hat{\sigma}^{(2)}_{3}$ operators for photon-$1$ and photon-$2$ separately. For photon-$2$, a measurement can be performed simultaneously or in any time order.  The measurement results from distant locations can be combined together at a later time in the future in frame $S$. By multiplying the measurement outcomes of each experiment and repeating the experiment many times the expectation value of the measured operators can be evaluated. A decision about which operator to be measured can be taken even after the emission of photons without any prior knowledge about their quantum state. A different but known selection of measurement basis for one entangled photon can steer the distant photon quantum state \cite{steering_doherty, steering_doherty2, steering_Reid1, steering_pryde, steering_zeilinger}. The polarization entangled state is also an eigen state of an operator $\hat{\sigma}^{(1)}_{1}\otimes\hat{\sigma}^{(2)}_{1}$ such that $\hat{\sigma}^{(1)}_{1}\otimes\hat{\sigma}^{(2)}_{1}|\Psi;t\rangle_{E}=-|\Psi;t\rangle_{E}$. However, an expectation value of measurements on any single photon of these operators $\langle\hat{\sigma}^{(j)}_{1}\rangle=\langle\hat{\sigma}^{(j)}_{2}\rangle=\langle\hat{\sigma}^{(j)}_{3}\rangle$ is zero. A particular combination of operators corresponding to measurements on both photons violates a Bell's type of inequality  \cite{chsh_th}, which is $|\langle \hat{a}_{1} \hat{a}_{2}\rangle+\langle \hat{a}_{1} \hat{b}_{2}\rangle+\langle \hat{b}_{1} \hat{a}_{2}\rangle-\langle \hat{b}_{1} \hat{b}_{2}\rangle| \leq2$, where $\hat{a}_{1}=\hat{\sigma}^{(1)}_{3}$, $\hat{b}_{1}=\hat{\sigma}^{(1)}_{1}$ and $\hat{a}_{2}=-(\hat{\sigma}^{(2)}_{1}+\hat{\sigma}^{(2)}_{3})/\sqrt{2}$,  $\hat{b}_{2}= (\hat{\sigma}^{(2)}_{1}-\hat{\sigma}^{(2)}_{3})/\sqrt{2}$ are operators corresponding to measurements performed on photon-$1$ and photon-$2$, respectively. In frame $S'$, polarization entangled state is  $|\Psi';t'\rangle_{E}=(|\hat{z}'\rangle_{1}|\hat{y}'\rangle_{2}-|\hat{y}'\rangle_{1}|\hat{z}'\rangle_{2})/\sqrt{2}$ given in Eq.~\ref{eq4}. According to transformations described previously, corresponding operators in $S'$ are defined as $\hat{\sigma}^{(j)}_{1}\rightarrow\hat{\sigma}'^{(j)}_{1}\doteq(|\hat{z}'\rangle\langle \hat{y}'|+|\hat{y}'\rangle\langle \hat{z}'|)_{j}$, $\hat{\sigma}^{(j)}_{2}\rightarrow\hat{\sigma}'^{(j)}_{2}\doteq i(-|\hat{z}'\rangle\langle \hat{y}'|+|\hat{y}'\rangle\langle \hat{z}'|)_{j}$
$\hat{\sigma}^{(j)}_{3}\rightarrow \hat{\sigma}'^{(j)}_{3}\doteq(|\hat{z}'\rangle\langle \hat{z}'|-|\hat{y}'\rangle\langle \hat{y}'|)_{j}$ and $\hat{I}^{(j)}=\hat{I}'^{(j)}$. If measurements are performed in frame $S'$ instead of frame $S$ by stationary detectors in frame $S'$, then a following Bell's inequality is violated that is $|\langle \hat{a}'_{1} \hat{a}'_{2}\rangle+\langle \hat{a}'_{1} \hat{b}'_{2}\rangle+\langle \hat{b}'_{1} \hat{a}'_{2}\rangle-\langle \hat{b}'_{1} \hat{b}'_{2}\rangle| \leq2$, where $\hat{a}'_{1}=\hat{\sigma}'^{(1)}_{3}$, $\hat{b}'_{1}=\hat{\sigma}'^{(1)}_{1}$ and $\hat{a}'_{2}=-(\hat{\sigma}'^{(2)}_{1}+\hat{\sigma}'^{(2)}_{3})/\sqrt{2}$,  $\hat{b}'_{2}= (\hat{\sigma}'^{(2)}_{1}-\hat{\sigma}'^{(2)}_{3})/\sqrt{2}$ are operators in frame $S'$ corresponding to operators $\hat{a}_{1}$, $\hat{b}_{1}$, $\hat{a}_{2}$,  $\hat{b}_{2}$ in frame $S$. These operators in frame $S'$ act on the frame transformed quantum entangled state $|\Psi';t'\rangle_{E}$ and Bell's inequality is violated in all inertial frames of reference.

\section{Nonlocal action in the past}
\begin{figure}[ht]
\centering
\includegraphics[scale=0.57]{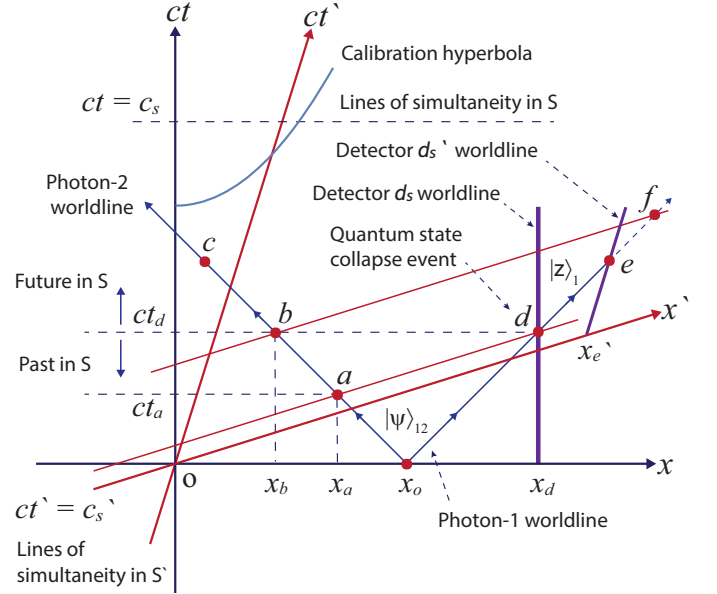}
\caption{\label{fig1} \emph{Spacetime diagrams representing effect of a measurement on a quantum entangled state of photons in frames $S$ and $S'$ connected by the Lorentz transformations. Photon-$1$ quantum state is measured by a stationary detector $d_{s}$ in frame $S$ at a spacetime location \textbf{d} and  quantum state of photon-$2$ is determined simultaneously at a spacetime location \textbf{b} at time $t_{d}$. However, in frame $S'$, events \textbf{d} and \textbf{b} are not simultaneous but an event \textbf{d} is simultaneous with a spacetime location \textbf{a}, where \textbf{a} is located in the past of \textbf{d} and \textbf{b} in frame $S$. In order to be consistent with the measurement outcomes in both frames the quantum state of photon-$2$ has to collapse in the past provided one photon is in the future of quantum state collapse event in frame $S$.}}
\end{figure}

Assume that quantum  entangled photons produced around a spacetime location ($x_{o}, 0$) in frame $S$ with total quantum state Eq.~\ref{eq3} are propagating in opposite direction along $x$-axis. Consider a spacetime diagram representing events \emph{w.r.t.} frames $S$ and $S'$  as shown in Fig.~\ref{fig1}, where planes $x$-$ct$ of frame $S$ and $x'$-$ct'$ of frame $S'$ are considered since other two spatial dimensions are unaffected by the Lorentz transformations under consideration.  A coordinate transformation from frame $S$ to frame $S'$ follows hyperbolic geometry, which is a result of invariance of quantity $x^{2}-(ct)^{2}= x'^{2}-(ct')^{2}$ since speed of light is invariant in all inertial frames of reference.  Space and time coordinate axes are calibrated by using this invariance.  Propagation of  photons is represented by worldlines. Interaction of a photon with a detector is represented by intersection of their respective worldlines.
Polarization quantum state of photon-$1$ is measured by a polarization sensitive single photon detector $d_{s}$, which is at rest in frame $S$ and positioned at $x_{d}$, where $x_{d}>x_{o}$.  Detector is a quantum state measurement device and it detects photon-$1$ in $|\hat{y}\rangle_{1}$ polarization quantum state when it reaches at detector at time $t_{d} $. However, detector is transparent to an orthogonal quantum state $|\hat{z}\rangle_{1}$. Detector can be rotated to detect any other component of linear polarization quantum state and to transmit its orthogonal component without detection. Therefore, if photon-$1$ is measured in quantum state $|\hat{y}\rangle_{1}$ by the detector then an output signal is produced which signifies a measurement outcome to be $|\hat{y}\rangle_{1}$ and polarization entangled state $|\Psi;t\rangle_{E}$ is collapsed on to $|\hat{y}\rangle_{1}|\hat{z}\rangle_{2}$. If on the other hand, detector does not produce any output signal then photon is passed through it and polarization entangled state is collapsed on to a quantum state $|\hat{z}\rangle_{1}|\hat{y}\rangle_{2}$.  In this way,  detector performs a polarization selective measurement. The collapse of a polarization entangled state happens when photon-$1$ interacts with detector \emph{i.e.} when $\psi_{1}(x_{1},t)$ overlaps with detector, regardless of its output state since detector is assumed to be one hundred percent efficient. Suppose, photon-$1$ is measured by a stationary detector $d_{s}$ in a very small spacetime region around a spacetime location $(x_{d}, ct_{d})$ in frame $S$ represented by a point \textbf{d} as shown in Fig.~\ref{fig1} .
      Polarization quantum state of any single photon prior to quantum state collapse is undefined even in principle due to their quantum entanglement and considered to be completely random, which leads to  $\langle\hat{\sigma}^{(j)}_{1}\rangle=\langle\hat{\sigma}^{(j)}_{2}\rangle=\langle\hat{\sigma}^{(j)}_{3}\rangle=0$. Immediately after the quantum state collapse at a spacetime point \textbf{d}, polarization quantum state of photon-$2$, which can be propagating at large distance from photon-$1$, is precisely determined from the measurement outcome in frame $S$ without making any interaction with photon-$2$. A spacetime point \textbf{d} is defined as the quantum state collapse event. Immediately after the polarization measurement of photon-$1$ at time $t_{d}$, the location of photon-$2$ is given by its spatial wavefunction $\psi_{2}(x_{2},t_{d})$.

Quantum state collapse happens along a line of simultaneity in frame $S$ that is $t=t_{d}$ passing through the quantum state collapse event \textbf{d}.
Prior to the quantum state collapse that is $0<t<t_{d}$, which is past in frame $S$, photons are polarization entangled where total quantum state is given by Eq.~\ref{eq3}. For $t>t_{d}$, which is future in frame $S$,  quantum state of photons is collapsed and unentangled. A question arises here, has the collapse of quantum entangled state influenced quantum state of the distant photon in its past in frame $S$? This question is significant since past and future events can coexist for a fast moving inertial observer. Therefore, how to determine if quantum entangled state has collapsed in the past. Interestingly, this is possible to know as explained further.

 In frame $S$, quantum entangled state collapse happens around a point $\textbf{d}$ and the nonlocal action happens simultaneously around a point
$\textbf{b}$, which is an intersection of wordline of photon-$2$ and a line of simultaneity $ct=c_{s}$ in frame $S$, where a constant $c_{s}=ct_{d}$.
  The action of a measurement  determines the quantum state of separated photons simultaneously at all space point at time $t_{d}$. Furthermore, a measurement can be performed in a diagonal basis, $|\hat{d}^{+}\rangle_{j}=(|\hat{z}\rangle_{j}+|\hat{y}\rangle_{j})/\sqrt{2}$, $|\hat{d}^{-}\rangle_{j}=(|\hat{z}\rangle_{j}-|\hat{y}\rangle_{j})/\sqrt{2}$, by rotating detector $d_{s}$, which is equivalent to measuring an operator $\hat{\sigma}^{(1)}_{1}$. If detector $d_{s}$ transmits $|\hat{d}^{+}\rangle_{1}$ and its output state remains low then quantum state collapse at a point $\textbf{d}$ simultaneously determines  quantum state of photon-$2$ to be $|\hat{d}^{-}\rangle_{2}$ at a point $\textbf{b}$ without making any measurement on it. Where a choice of measurement basis is random and a decision can be made just prior to interaction of photon-$1$ with detector $d_{s}$ that is at $t\lesssim t_{d}$, which can be much later than time of emission of photons \emph{i.e.} $t_{d}\gg0$.
Therefore, after the quantum state collapse each photon has a well defined polarization state. However, simultaneous events of frame $S$ are not simultaneous \emph{w.r.t.} another inertial frames of reference moving at relativistic speed.  Lines of simultaneity where time coordinate is kept constant are different \emph{i.e.} $ct=c_{s}$ in frame $S$ and $ct'=c'_{s}$ in frame $S'$, where $c_{s}$ and $c'_{s}$ are  constant. Therefore,  quantum state collapse event points \textbf{d} and \textbf{b}, which are simultaneous in frame $S$ are not simultaneous in frame $S'$.

In frame $S'$, polarization entangled state of photons prior to any measurement is given by Eq.~\ref{eq4} which is  $|\Psi'; t'\rangle_{E}=\frac{1}{\sqrt{2}}(|\hat{z}'\rangle_{1}|\hat{y}'\rangle_{2}-|\hat{y}'\rangle_{1}|\hat{z}'\rangle_{2})$. Polarization entangled state is also time independent in frame $S'$ and time argument is retained to show its simultaneity in frame $S'$. Quantum state collapse in frame $S$ at $\textbf{d}$ can be confirmed in frame $S'$. Consider a stationary detector $d'_{s}$ in frame $S'$ at a location $x'_{e}$ such that it detects photon-$1$ at a spacetime point $\textbf{e}$ after  quantum state collapse by detector $d_{s}$ at a point $\textbf{d}$.  Detector $d'_{s}$ is also a polarization selective detector  which produces an output signal if photon-$1$ is detected in quantum state $|\hat{z}'\rangle_{1}$, which is the undetected  state $|\hat{z}\rangle_{1}$ by detector $d_{s}$ in frame $S$.  Detectors $d_{s}$ and $d'_{s}$ are aligned such that the transmitted quantum state of photon-$1$ by detector $d_{s}$ is completely detected by detector $d'_{s}$ and its orthogonal component is undetected and transmitted. Detectors can be inserted in path of photon-$1$ from a plane orthogonal to the direction of velocity of frame $S'$ to circumvent their collision.

Assume detector $d_{s}$ collapsed  $|\Psi;t\rangle_{E}$ on to  $|\hat{z}\rangle_{1}|\hat{y}\rangle_{2}$. Therefore, detector $d'_{s}$ definitely detects photon-$1$. If detector $d'_{s}$ is not placed in path of photon-$1$ then photon-$1$ intersects with a line of simultaneity of frame $S'$, at a point $\textbf{f}$, passing through a point $\textbf{b}$. If detector location $x'_{e}$ is chosen such that its worldline intersects with worldline of photon-$1$ between  points $\textbf{d}$
 and $\textbf{f}$ at an arbitrary point $\textbf{e}$ then a corresponding simultaneous location of photon-$2$ is located on the worldline of photon-$2$ between points $\textbf{a}$
and $\textbf{b}$ as shown in Fig.~\ref{fig1}. However, this region corresponds to past in frame $S$, where location of photon-$2$ is given by its position wavefunction $\psi'_{2}(x'_{2},t')$. In the past, photon-$2$ was quantum entangled with photon-$1$ but in frame $S'$ the quantum state of photon-$1$ is definitely known to be $|\hat{z}'\rangle_{1}$ after the quantum state collapse event $\textbf{d}$ for $t>t_{d}$ even if detector $d'_{s}$ is not placed. Since photons were quantum entangled in all inertial frames therefore, from the invariance of quantum entanglement shown before, the quantum state of both photons after the collapse in $S'$ should be $|\hat{z}'\rangle_{1}|\hat{y}'\rangle_{2}$ in order to be consistent with the corresponding collapsed quantum state $|\hat{z}\rangle_{1}|\hat{y}\rangle_{2}$ in frame $S$. If on the other hand, quantum state is collapsed on to $|\hat{y}\rangle_{1}|\hat{z}\rangle_{2}$ in frame $S$ then the collapsed quantum state in frame $S'$ is $|\hat{y}'\rangle_{1}|z'\rangle_{2}$. This is consistent only if quantum state of photon-$2$ is nonlocally collapsed in frame $S$ not only at a point $\textbf{b}$ but the collapse also happened in the past. Otherwise, quantum state after the quantum state collapse will be different than the respective component of quantum entangled state in different inertial frames of reference.  Lorentz transformed measurement outcome must agree in all inertial frames of reference along their line of simultaneity.
Furthermore, the essence of quantum entanglement is strengthened by random choice of measurement basis. If on the other hand, detector $d_{s}$ is aligned to measure $\hat{\sigma}^{(1)}_{1}$ for photon-$1$ after emission of photons and $|d^{+}\rangle_{1}$ is measured then  polarization entangled state is collapsed on to $|\hat{d}^{+}\rangle_{1}\hat{d}^{-}\rangle_{2}$ at simultaneous points $\textbf{d}$ and $\textbf{b}$ in frame $S$. In frame $S'$, polarization entangled state is collapsed on to $|\hat{d}'^{+}\rangle_{1}\hat{d}'^{-}\rangle_{2}$, where $|\hat{d}'^{-}\rangle_{2}$ is a quantum state in the past \emph{w.r.t.} the quantum state collapse event in frame $S$. Therefore, nonlocal action at a distance also acts in the past.

Quantum state collapse outcome must be consistent in all inertial frames of reference. For another inertial frame $S''$, moving with negative velocity along $x$-axis \emph{w.r.t.} frame $S$, the time order of events \textbf{d} and \textbf{b} is reversed since they are separated by a spacelike interval. An event on photon-$2$ worldline at a point \textbf{c} happens before the quantum state collapse event \textbf{d} in frame $S''$ \emph{i.e.} photon-$1$ has not been measured along a line of simultaneity passing through \textbf{c} in frame $S''$. An event \textbf{c} can correspond to quantum state analysis of photon-$2$ by a polarization selective photon detector. A line of simultaneity in frame $S''$ passing through \textbf{c} is intersecting with the worldline of photon-$1$ between \textbf{d} and point of emission of photons ($x_{o},0$). However, photon-$2$ is existing in the future in frame $S$ \emph{i.e.} it has crossed the line of simultaneity $t=t_{d}$, a time when quantum state collapse happened in frame $S$. In the future, photon-$2$ polarization quantum state is well defined therefore, polarization quantum state of photon-$1$, which is still in the past in frame $S$, has to be well defined even when an event corresponding to its interaction with detector $d_{s}$  occurs later in frame $S''$.  The Lorentz transformed quantum state collapse outcomes must be consistent in all inertial frames of reference. This analysis leads to a principle, which states that if an entangled quantum state of separated photons is collapsed in one inertial frame of reference then it is also collapsed in all inertial frames of reference along their respective lines of simultaneity provided any one photon is in the future of the quantum state collapse event in frame $S$. In addition, the result of this paper resolves the paradoxes related to ambiguity of quantum state collapse of separated particles for different inertial observers as described in References \cite{ayp1, wayne1}.

Spacetime locations of events in frames $S$ and $S'$ are evaluated by the Lorentz transformations. In frame $S$, spacetime coordinates of quantum state collapse event $\textbf{d}$ are ($x_{d}, ct_{d}$) and of point $\textbf{b}$ are ($x_{b}, ct_{d}$), where $x_{d}=x_{o}+ct_{d}$ and $x_{b}=x_{o}-ct_{d}$, these points are located on a line of simultaneity $ct=ct_{d}$ of frame $S$. In frame $S'$, spacetime coordinates of quantum state collapse event $\textbf{d}$ are $x'_{d}=\gamma (x_{d}-vt_{d})$,  $t'_{d}=\gamma(t_{d}-\frac{v}{c^{2}} x_{d})$ and of point $\textbf{b}$ are $x'_{b}=\gamma (x_{b}-vt_{d})$,  $t'_{b}=\gamma(t_{d}-\frac{v}{c^{2}} x_{b})$. A space interval between $\textbf{a}$ and $\textbf{d}$ in frame $S'$ is $(x'_{d}-x'_{a})^{2}=(x_{d}-x_{a})^{2}-c^{2}(t_{d}-t_{a})^{2}$. Since points $\textbf{d}$ and $\textbf{a}$ are located on the line of simultaneity $ct'=ct'_{d}$ in frame $S'$ therefore, their time coordinates $t'_{d}=\gamma(t_{d}-\frac{v}{c^{2}} x_{d})$ and $t'_{d}=\gamma(t_{a}-\frac{v}{c^{2}} x_{a})$ are same. Therefore, time interval which is defined as past in frame $S$ is $t_{d}-t_{a}=\frac{v}{c^{2}}(x_{d}-x_{a})$. For an arbitrary velocity $v< c$, maximum time interval of the past is time difference between the quantum state collapse and emission events of photons in frame $S$.

\section{Conclusion}
In conclusion, it is shown that the nonlocal collapse of an entangled quantum state happens not only in the present and proceeds in the future but it also happens in the past of the distant unmeasured photon in an inertial frame of reference in which a measurement is performed only on single photon by a stationary quantum state measurement device. In a relativistically moving inertial frame of reference past and future events can coexist. Since polarization entangled state under consideration is Lorentz invariant therefore, the Lorentz transformed quantum state collapse outcome must correlate in all inertial frames of reference along their respective line of simultaneity.   It is shown from the relative simultaneity, the validity of Bell's theorem and the invariance of quantum entanglement for all inertial observers that quantum state of the distant photon must correlate in its past also with the measurement outcome of the measured photon in all inertial frames of reference. Therefore, the  quantum state of the distant photon is unentangled from the quantum state of the measured photon even in the past of the distant photon.  Therefore, it is concluded that the nonlocal action at a distance also acts in the past.

\subsection*{Acknowledgement}
Mandip Singh (MS) acknowledges research funding by the Department of Science and Technology, for project No. Q.101., DST/ICPS/QuST/Theme-1/2019 (General).

\bibliography{sample}

\end{document}